\documentclass[a4paper,pdflatex]{article}
\usepackage{graphicx,amsmath,amsfonts}

\def \ba{\begin{eqnarray}}\def\ea{\end{eqnarray}}
\def\bc{\begin{center}}\def\ec{\end{center}}
\def\nn{\nonumber\\}

\onecolumn

\title{\bf  An improvement of the Moli\`{e}re–-Fano
multiple \\scattering theory}
\author{\bf A. Tarasov and O. Voskresenskaya\footnote{On leave of
absence from Siberian Physical Technical Institute. Electronic
address: voskr@jinr.ru}}
\date{}
\begin{document}

\maketitle
 \bc
Joint Institute for Nuclear Research, 141980 Dubna, Russia \ec

\begin{abstract}
\noindent In the framework of unitary Glauber approximation for
particle-atom scattering, we develop the general formalism of the
Moli\`{e}re--Fano multiple scattering theory (M--F theory) on the
basis of reconstruction of the generalized optical theorem in it. We
present rigorous relations  between some exact and first-order
parameters of the Moli\`{e}re multiple scattering theory, instead of
the approximate one obtained in the original paper by Moli\`{e}re.
We consider the relative unitarity corrections and the Coulomb
corrections to the quantities of the M--F theory. Also, we  examine
their $Z$ dependence in the range of nuclear charge from $Z=4$ to
$Z=92$. Additionally, we show the difference between our results and
those of Moli\`{e}re over this  range of $Z$.

\end{abstract}

\section{Introduction}

The Moli\`{e}re--Fano multiple scattering theory of charged
particles [1--3] is the most used tool for taking into account the
multiple scattering effects in experimental data processing. The
experiment DIRAC \cite{Dirac05}and many others (\cite{stand}, MuScat
\cite{others1}, MUCOOL \cite{others2} experiments, etc.) face the
problem of excluding the multiple scattering effects in matter from
obtained data.

The standard theory of multiple scattering
\cite{Dirac05,stand,others1} proposed by Moli\`{e}re \cite{M47,M55}
and Fano \cite{Fano} and some its modifications [6--11] are used for
this aim. The modifications, developed in [6--8], are motivated by
experiments \cite{others1,others2};  they are connected with
including analogues of the Fano corrections in the Moli\`{e}re
theory and determining their range of applicability [6--9]. In
\cite{Yur} a modified transport equation is presented whose solution
is applicable over the range of angles, from $0$ to $180^{\circ}$.
In \cite{MIFI} results of experiments \cite{ex} are qualitatively
explained within the framework of the theory allowing for pair
correlations in the spatial distribution of scatterers.

Estimation of the theory accuracy is of particular importance for
the DIRAC experiment because it's high angular resolution. One
possible source of the M--F theory inaccuracy is use in [1--3] an
approximate expression for the target-elastic particle-atom
scattering amplitude which violates the generalized optical theorem
\ba
 \Im
f_{el}(0)=\frac{k}{4\pi}\sigma_{tot} =
\frac{k}{4\pi}(\sigma_{el}+\sigma_{in})
\ea
or, in other words, unitarity condition. Another possible source of
inaccuracy is using in calculations an approximate relation for the
exact and the Born values of the screening angle
($\chi_a^{\,\prime}$)
\ba\label{m}\chi_a^{\,\prime}\approx\big(\chi_a^{\,\prime}\big)^{\scriptscriptstyle
B}\sqrt{1+3.34\left(Z\alpha\right)^2} \ea
obtained in the original paper by Moli\`{e}re \cite{M47}. Therefore,
the problem of estimating the M--F theory accuracy  and its
improvement becomes important.

In this work, we obtain the relative unitarity corrections to the
parameters of the Moli\`{e}re--Fano theory  both analytically and
numerically resulting from a reconstruction of the unitarity in the
particle-atom scattering theory, and we found that they are of an
order of $Z\alpha^2$. Also, we consider the analytical and numerical
results for the Coulomb corrections to the parameters of the
Moli\`{e}re theory, and we show that these corrections can be
numerically large, e.g., about $40\div 50\%$ for $Z=92$.
Additionally, we demonstrate the difference between our results and
those of Moli\`{e}re over the range $4\leq Z \leq 92$.

%and have concluded that
%for heavy atoms of the target material they become significant.

The paper is organized as follows. In Section 2, we consider the
approximations of the M--F theory. In Section 3, we obtain the
analytical and numerical results for the unitarity corrections and
the Coulomb corrections to the parameters of the  M--F theory. In
Conclusion, we briefly summarize our results.

\section{Approximations of the M--F theory}

\subsection{ Small-angle approximation}

Let all scattering angles are small $\theta \ll 1$ so that $\sin
\theta\sim\theta$, and the scattering problem is equivalent to
diffusion in the plane of $\theta$. Now let $\sigma^{}_{el}(\chi)$ be the
elastic differential cross section for the single scattering into
the angular interval $\vec\chi =\vec\theta-\vec\theta^\prime$, and
 $W_{\scriptscriptstyle M}(\theta,t)\theta d\theta$ is the number of scattered
 particles in the interval $d\theta$ after traversing a target
 thickness $t$. Then, within the small-angle approximation, the
 transport equation for the distribution function $W_{\scriptscriptstyle
M}(\theta,t)$ reads
\ba\label{kinet} \frac{\partial W_{\scriptscriptstyle
M}(\theta,t)}{\partial t}= -n_0 W_{\scriptscriptstyle
M}(\theta,t)\int \sigma^{}_{el}(\chi)d^2\chi +n_0 \int W_{\scriptscriptstyle
M}(\vec\theta-\vec \chi,t)\sigma^{}_{el}(\chi)d^2\chi,\ea
where $n_0$ is the density of the scattering centers per unit
volume, $d^2\chi=\chi d\chi d\phi/(2\pi)$, and $\phi$ denotes the
azimuthal angle of the vector $\vec{\chi}=(\theta,\phi)$.

Introducing the Bessel transformation of distribution
\ba\label{eq3} g(\eta,t)=\int\limits_0^\infty \theta
J_0(\eta\theta)W_{\scriptscriptstyle
M}(\theta,t)d\theta\ ,  \\
\label{W}W_{\scriptscriptstyle M}(\theta,t)=\int\limits_0^\infty
\eta J_0(\eta\theta)g(\eta,t)d\eta\ , \ea
and using the folding theorem (see details in Appendix), we obtain
the transport equation for the Bessel-transformed  function
$g(\eta,t)$:
\ba\label{eq5} \frac{\partial g(\eta,t)}{\partial t}=-n_0\,
g(\eta,t)\int\limits_0^\infty\sigma_{el}(\chi) \chi
d\chi[1-J_0(\eta\chi)]\  ,\ea
whose solution is
\ba \label{g(eta)}g(\eta,t)=\exp\left\{N(\eta,t)-N_0(0,t)\right\}\
,\ea
\ba N(\eta,t)=n_0\, t\int\sigma_{el}(\chi)\chi d\chi J_0(\eta\chi)\
. \ea

Inserting this solution back in (\ref{W}), we get
\ba\label{com} W_{\scriptscriptstyle
M}(\theta,t)=\int\limits_0^\infty\eta d\eta
J_0(\eta\theta)\exp\left\{-n_0\, t \,\int\limits_0^\infty
\sigma_{el}(\chi)\chi d\chi\left[1-J_0(\eta\chi)\right]\right\}. \ea
This equation  is exact for any scattering law, provided only the
angles are small compared with a radian.

\subsection{Approximate solution of the transport equation}

For the screening potential, the scattering cross section reads
\ba
    \sigma_{el}(\chi) &=& \frac{4 Z^2 e^4}{(v p)^2
    (\chi^2+\chi_0^2)^2},
\ea
\ba \chi^{}_0=\frac{\lambda}{a}\,,\quad \lambda=\frac{\hbar}{m
v}\,,\quad a=0.885\,a^{}_0Z^{-1/3}\,,\quad
a^{}_0=\frac{\hbar}{p\alpha}, \ea
where $a$ is the Fermi radius of the atom, $a^{}_0$ presents the
Bohr radius of the particle, $p=mv$ and $v$ are the incident
particle momentum and the particle velocity in the laboratory frame,
correspondingly, and $\alpha=1/137$ denotes the fine structure
constant.

Let us write
\ba n^{}_0 t\,\sigma_{el}(\chi)\chi d\chi=2\chi_c^2\chi d\chi
\frac{q_{el}(\chi)}{\chi^4},\qquad 2\chi_c^2= 4\pi n^{}_0 t
\frac{Z^2e^4}{(p v)^2},\ea
where $q_{el}(\chi)$ is the ratio of the actual scattering cross
section to the Rutherford one, and $\chi_c$ is the so-called
`characteristic angle'
\ba\label{char} \chi_c^2=4\pi n_0t\left(\frac{Z\alpha}{\beta p}
\right)^2, \ea
whose physical meaning is that the probability of scattering on the
angles exceeding $\chi^{}_c$ is unity.

In terms of $\chi^{}_c$ and $q_{el}$, the solution of \eqref{eq5}
becomes
\ba -\ln
g(\eta,t)=2\,\chi_c^2\int\limits_0^\infty\frac{d\chi}{\chi^3}\,q_{el}(\chi)\left[1-J_0(\chi
\eta)\right]. \ea
Estimating the value of the latter integral and introducing the
notion of the screening angle $\chi^{}_a$
\ba -\ln\chi^{}_a=\lim_{\varsigma \to
\infty}\left[\int\limits_0^\varsigma\frac{d\chi}{\chi}\,q_{el}(\chi)+
\frac{1}{2}-\ln \varsigma\right], \ea
where $\chi^{}_0\ll \varsigma\ll\chi^{}_c$, we obtain for the
Bessel-transformed distribution function $g(\eta,t)$ the expression
\ba -\ln
g(\eta,t)=\frac{1}{2}(\chi^{}_c\eta)^2\left[-\ln(\chi^{}_a\eta)+\frac{1}{2}+\ln
 2-C_{\scriptscriptstyle
E}\right]. \ea
Here, $C_{\scriptscriptstyle E}=0.5772\ldots$ is the Euler constant.
Then introducing the new variables $y=\chi^{}_c\eta$ and
$\theta/\chi_c=u$, as a result, we get
\ba -\ln
g(\eta,t)=\frac{1}{4}y^2\left[b_{el}-\ln\left(\frac{1}{4}y^2\right)\right], \nn \\
b_{el}=\ln\frac{\chi_c^2}{\chi_a^2}+1-2C_{\scriptscriptstyle
E}\equiv\ln\frac{\chi_c^2}{(\chi'_a)^2}, \ea
and Moli\`{e}re's transformed equation becomes
\ba\label{simpl} W_{\scriptscriptstyle M}(\theta)\theta d\theta = u
du\int\limits_0^{\infty}y dy J_0(u y)\exp\left\{
-\frac{y^2}{4}\left[b^{}_{el}-\ln\left(\frac{y^2}{4}\right)\right]\right\}.
\ea

This rather simple formula permits one to develop an iteration
procedure for $W(\theta)$. Really, putting
\ba\label{B} b_{el}=B-\ln B,\ea
we can write the angular distribution function  as
\ba\label{exp2} W_{\scriptscriptstyle M}(\theta,B) &=&\frac{1}{
\overline{\theta^{\,2}}}\int\limits_0^{\infty}y dy J_0 (\theta
y)e^{-y^2/4}\exp\left[\frac{y^2}{4B}\ln\left(\frac{y^2}{4}\right)\right],
\ea
with $\overline{\theta^{\,2}} = \chi_c^2 B$. Introducing the
variable $x=u^2/B$, one can obtain the expansion of the distribution
function in a power series in $1/B$:
\ba\label{ex} W_{\scriptscriptstyle M}(\theta,B)\theta d\theta&=&
\frac{1}{\overline{\theta^{\,2}}}\theta
d\theta\left[W^{(0)}(x)+\frac{1}{B}W^{(1)}(x)+
\frac{1}{B^2}W^{(2)}(x)+\ldots\right],\ea
in  which
 \ba\label{exp}
W^{(0)}(x)=\frac{2}{\overline{\theta^{\,2}}}\exp\left(\!\!-
\frac{\theta^2}{\overline{\theta^{\,2}}}\right),\quad\ldots \;,\quad
W^{(n)}(x)=\frac{1}{\overline{\theta^{\,2}}}\int\limits_0^{\infty}y
dy J_0 \left(\frac{\theta}{\bar\theta}\,
y\right)e^{-y^2/4}\left[\frac{y^2}{4}\ln\left(\frac{y^2}{4}\right)\right]^n.
 \ea

The result of numerical integration $W^{(n)}(x)$ was obtained in
papers by Moli\`{e}re, Bethe and Scott \cite{M47,Bethe,Scott49}. In
practice, the value of the solution of the transcendental equation
\eqref{B} is large enough to provide the convergence of the
expansion series, i.e., $4.5 \leq B\leq 20$.

\subsection{Born approximation}

On the one hand, Moli\`ere writes the elastic Born cross section for
fast charged particle Coulomb scattering as follows:
\ba\label{q} \sigma_{el}^{\scriptscriptstyle
B}(\chi)=\sigma_{el}^{\scriptscriptstyle
R}(\chi)\,\,q_{el}^{\scriptscriptstyle B}(\chi).\ea

For angles $\chi$ small compared with a radian, the Rutherford cross
section has a simple approximation, and (\ref{q}) yields
\ba\label{sigm}\sigma_{el}^{\scriptscriptstyle B}(\chi)&=&
\frac{\chi_c^2}{4\pi n_0 t
(1-\cos\chi)^2}\,\,q_{el}^{\scriptscriptstyle B}(\chi)\\\label{sig}
&\approx& \frac{\chi_c^2}{\pi n_0
t\,\chi^4}\,\,q_{el}^{\scriptscriptstyle B}(\chi). \ea

Moli\`ere represents  the Born screening angle
$\chi_a^{\scriptscriptstyle B}$ via $q_{el}^{\,\scriptscriptstyle
B}(\chi)$ by the equation
\ba\label{def} -\ln\big(\chi^{\scriptscriptstyle B}_a\big)
&=&\lim\limits_{\varsigma\rightarrow
\infty}\left[\int\limits_0^\varsigma\frac{q_{el}^{\,\scriptscriptstyle
B}(\chi)d \chi}{\chi}+\frac{1}{2}-\ln \varsigma\right]
 \ea
with
\ba\label{zeta}\chi_0\ll \varsigma\ll 1/\eta\sim \chi_c\ea
and $\chi_0\sim m_e\alpha Z^{1/3}/p$.

When the Born parameter $a$ is equal to zero, \eqref{def} can be
evaluated directly using the facts that $q(0)=0$ and
$\lim\limits_{\varsigma\rightarrow \infty}q(\varsigma)=1$. Then  the
following approximation  can be obtained for
$\big(\chi_a^{\,\prime}\big)^{\scriptscriptstyle B}$:
\ba \label{theta_a}\big(\chi_a^{\,\prime}\big)^{\scriptscriptstyle
B}=\left[\exp(C_{\scriptscriptstyle
E}-0.5)\right]\,\frac{\lambda}{p}\,\,1.065=\sqrt{1.174}\,\,\,\chi_0\,\sqrt{1.13},\ea
where  $\lambda=m_e \alpha Z^{1/3}/0.885$.

On the other hand, Moli\`ere writes the non-relativistic Born cross
section in the form
\ba \label{sigma} \sigma_{el}^{\scriptscriptstyle B}(\chi)=
k^2\left\vert\int\limits_{0}^{\infty}\rho\, d\rho
J_0\left(2k\rho\,\sin\frac{\chi}{2}\right)\Phi_{\scriptscriptstyle
M}^{\scriptscriptstyle B}(\vec \rho)\right\vert^2, \ea
in which the Born phase shift is given by
\ba\label{Phi_B} \Phi_{\scriptscriptstyle M}^{\scriptscriptstyle
B}(\vec \rho)=
-\frac{2}{v}\int\limits_{\rho}^{\infty}\frac{U_{\lambda}(r)rdr}{\sqrt{r^2-\rho^2}}=
-\frac{1}{v}\int\limits_{-\infty}^{\infty}U_{\lambda}\left(r=\sqrt{\rho^2+z^2}
\right)dz \ea
in units of $\hbar=c=1$ Here, $k$ is the wave number of the incident
particle, the variable $\rho$ corresponds to the impact parameter of
the collision,
 and $U_{\lambda}(r)$ is a screened Coulomb potential of the target
atom.

The Born target-elastic single cross section satisfies the following
relations:
\ba \frac{d\sigma^B_{el}}{d\Omega}=\vert f_{el}(\theta)\vert^2,\ea
 \ba \Im f_{el}(0)&=&
\frac{k}{4\pi}\sigma^B_{el}\neq \frac{k}{4\pi} \sigma^B_{tot}. \ea
The Born approximation result for
 the target-elastic scattering amplitude $f_{el}$ with the momentum transfer
$q=k\theta$  reads
\ba f_{el}(\theta)=ik \int\limits_0^{\infty} J_0(\rho\, q)[1-
e^{i\Phi_{\scriptscriptstyle M}^{}(\vec \rho)}]\rho \,d\rho. \ea

\subsection{Approximate relation between the quantities $\chi_a$ and
$\chi_a^{\scriptscriptstyle
B}$}

For actual determining the screening angle
\ba\label{de} -\ln\big(\chi^{}_a\big)
&=&\lim\limits_{\varsigma\rightarrow
\infty}\left[\int\limits_0^\varsigma\left(1-\frac{F_{\scriptscriptstyle
A}(p\chi)}{Z}\right)^2\frac{
d\chi}{\chi}+\frac{1}{2}-\ln \varsigma  \right]\\
\label{defin}&=&\lim\limits_{\varsigma\rightarrow
\infty}\left[\int\limits_0^\varsigma\frac{q_{el}^{}(\chi)d
\chi}{\chi}+\frac{1}{2}-\ln \varsigma\right]
 \ea
via the Thomas--Fermi form factor $F_{\scriptscriptstyle T-F}(q)$
\ba\label{F} F_{\scriptscriptstyle T-F}(q)^{\scriptscriptstyle
M}=\sum\limits_{i=1}^{3}\frac{c_i\lambda_i^2}{q^2+\lambda_i^2}\,,
\ea
where
$$c_1=0.35,\quad c_2=0.55,\quad c_3=0.10,$$
$$\lambda_1=0.30\lambda,\quad \lambda_2=4\lambda_1,\quad \lambda_3=5\lambda_2,$$
which does not make use of the Born approximation,  Moli\`{e}re uses
the WKB method.

He starts with the exact formulas for the WKB differential cross
section $\sigma_{el}(\chi)$ and the corresponding ratio
$q_{el}(\chi)$
\ba\label{WKB1} \sigma_{el}(\chi)=
k^2\left\vert\int\limits_{0}^{\infty}\rho \,d\rho \,J_0(k\chi
\rho)\bigg\{1-\exp\big[ i\Phi_{\scriptscriptstyle M}(\vec
\rho)\big]\bigg\}\right\vert^2,\ea
\ba\label{WKB1.5} q_{\,el}(\chi)=
\frac{(k\chi)^4}{4\,a^2}\left\vert\int\limits_{0}^{\infty}\rho\,
d\rho J_0(k\chi\rho)\bigg\{1-\exp \big[i\Phi_{\scriptscriptstyle
M}(\vec \rho)\big]\bigg\}\right\vert^2,\ea
but evaluates these quantities only approximately.

For estimation of \eqref{WKB1.5}, Moli\`{e}re uses the  Born shift
\eqref{Phi_B} with the potential
\ba\label{pot} U_{\lambda}(r)=\pm
Z\,\frac{\alpha}{r}\,\Lambda(\lambda r) \ea
and the Thomas--Fermi (T--F) screening function, which Moli\`{e}re
approximates by a sum of three exponentials
\ba\label{fit} \Lambda(\lambda r)&\simeq& 0.1e^{-6 \,\lambda r}+
0.55e^{-1.2 \,\lambda r} +0.35e^{-0.3 \,\lambda r}. \ea
Here, $r_{sc}=0.885/m_e \alpha Z^{1/3}$ is the T--F radius.

This is good only to terms of first order in the Born parameter $a$.
Neglecting terms of orders higher than $a^2$ in the obtained result,
he get the following approximate expression for $q_{el}(\chi)$:
\ba\label{WKB4} q^{}_{\,el}(\chi)\approx
1-\frac{8.85}{(\chi/\chi_0)^2} \bigg[1+ 2.303\,a^2\lg\frac{7.2\cdot
10^{-4}(\chi/\chi_0)^4} {\left(a^4+a^2/3+0.13\right)}\bigg],\ea
which is one of the basic expressions of the Moli\`{e}re theory.
Estimating $q_{el}(\chi)$ for different $a$ values, Moli\`{e}re
devised an interpolation scheme for $(\chi/\chi_0)^{2}$:
\ba (\chi/\chi_0)^{2}\approx A_q+a^2B_q. \ea
Finally, calculating the screening angle defined by
\ba\label{defex} -\ln\big(\chi_a\big)
=\frac{1}{2}-\ln\chi_0-\int\limits_0^1dq\ln\left(\frac{\chi}{\chi_0}\right)
\ea
and assuming a linear relation between $\chi_a^2$ and $a^2$, he get
the following interpolating formula for the screening angle:
\ba\label{interpol} \chi_a\approx\chi_0\sqrt{1.13+3.76\,a^2}.
%=\chi_a^{\scriptscriptstyle
%B}\sqrt{1+3.34\,a^2}.
\ea

\subsection{Fano approximation}

To estimate a contribution of incoherent scattering on atomic
electrons, the squared nuclear charge $Z^2$ is often replaced with
the sum of the squares of the nuclear and electronic charges
$Z(Z+1)$ \cite{M55,Fano,Bethe,Klch} in basic relations for
differential cross-section, some parameters of the theory, etc.

This procedure would be accurate if the single-scattering cross
sections were the same for nucleus and electrons of target atoms.
Besides, the actual cross sections are different at small and large
angles. Fano modified the multiple scattering theory  taking into
account above differences.

For this purpose, Fano separates the elastic and inelastic
contributions to the cross section \ba\label{fano}\sigma(\chi)=
\sigma_{el}(\chi)+ \sigma_{in}(\chi).\ea

For the inelastic components of the single scattering differential cross sections,
 the Fano approximation reads
\ba \frac{d\sigma_{in}}{d\Omega}= \frac{d\sigma^{\scriptscriptstyle
B}_{in}}{d\Omega}.\ea

Since the Born
single-scattering amplitudes are pure real, the generalized
optical theorem cannot be used to calculate the total cross section
in the framework of this approximation.

Fano sets the task of comparing the $\sigma^B_{in}(\chi)$
contribution to the exponent of the Goudsmit--Saunderson
distribution\footnote{ The Goudsmit--Saunderson theory is valid for
any angle, small or large, and do not assume any special form for
the differential scattering cross section.} \cite{GS}:
%$\chi$:
%
\ba \label{GS}W(\theta,t)=2\pi
\sum_l\left(l+\frac{1}{2}\right)P_l(\theta)\exp\left\{\!-n_0\,
t\!\int\! \sigma^{\scriptscriptstyle B}(\chi)\sin\chi
d\chi[1-P_l(\chi)] \right\}\!,\ea
where $P_l$ is the Legendre polynomial. If we replace  the sum over
$l$ in \eqref{GS} by an integral over $\eta$,
$\left(l+\frac{1}{2}\right)$ by $\eta$, $P_l$ by the well-known
formula $P_l(\theta)=J_0\big(\left(l+\frac{1}{2}\right)\theta\big)$,
and $\sin\chi$ by $\chi$, the expression \eqref{GS} goes over into
small-angle distribution  Moli\`ere's and Lewis' distribution
\eqref{com}.
%(9)

To achieve the mentioned goal in the small-angle approximation, we
determine the corresponding expressions for the inelastic cross
section
\ba\label{exact2}\sigma^{\scriptscriptstyle
B}_{in}(\chi)=\sigma^{\scriptscriptstyle
R}(\chi)\,q_{\,in}^{\scriptscriptstyle B}(\chi)
\label{sigmain}=\frac{\chi_c^2}{ 4\pi n_0
t\,\,Z\,(1-\cos\chi)^2}\,\,q_{\,in}^{\scriptscriptstyle B} (\chi)\ea
\ba\approx\frac{\chi_c^2}{ \pi n_0 t\,\,Z
\,\chi^4}\,\,q_{\,in}^{\scriptscriptstyle B}(\chi) \ea
and the `inelastic cut-off angle' $\chi_{in}^{\scriptscriptstyle
B}$
\ba -\ln\big(\chi^{\scriptscriptstyle
B}_{in}\big)=\lim\limits_{\varsigma\rightarrow
\infty}\left[\int\limits_0^\varsigma\frac{q_{\,in}^{\scriptscriptstyle
B}(\chi)d \chi}{\chi}+\frac{1}{2}-\ln \varsigma\right] \ea
similarly to \eqref{sigm} and \eqref{def}, in accordance with
\cite{Strig,Bethe}.

Then, using \eqref{sigm} and \eqref{sigmain}, we rewrite
\eqref{exp2} as follows:

\ba\label{expin} W_{\scriptscriptstyle M-F}(\Theta,B)
&=&\frac{1}{\overline{\theta^{\,2}}}\int\limits_0^{\infty}y dy J_0
(\Theta y)e^{-y^2/4}\exp\left(Y^{}_{el}+Y^{}_{in}\right) \ea
with
\ba Y^{}_{el}=\frac{y^2}{4B}\ln\left(\frac{y^2}{4}\right),\quad
Y^{}_{in}=\frac{2y^2}{(Z+1)B}\int\limits_\varsigma^\infty[1-J_0(\Theta)]\Theta^{-3}d\Theta,
\ea
where the parameter $B$ is defined by equation
\ba B-\ln B=b^{}_{el}+b^{}_{\,in}, \ea
in which
\ba b^{}_{\,el}=\ln\left(\frac{\chi_c}{\chi^{\scriptscriptstyle
B}_a}\right)^2+1-2C_{\scriptscriptstyle E},\quad
b^{}_{in}=\frac{1}{Z+1}\ln \left(\frac{\chi^{\scriptscriptstyle
B}_a}{\chi^{\scriptscriptstyle B}_{in}} \right)^2. \ea

Numerical estimation of the quantity
 $-u^{}_{in}{=}-\ln \left(\chi^{\scriptscriptstyle
B}_{in}\right)^2$ yields  $\left(-u^{}_{in}\right)_{\mathrm{T-F}}$
$=5.8$ for all $Z$ within the T--F model. This value  should not
vary greatly from one target material to another.

For sufficiently large angles, with use of exact Rutherford formulas
 \eqref{sigm} and \eqref{exact2}, the correct angular distribution
$W(\theta,t)$ can be estimated according to the formula \ba
W_{corr}(\theta,t)=W(\theta,t)
\left[\sigma^B_{exact}(\chi)/\sigma^{R}(\chi\right)],\ea as
suggested Bethe and Fano \cite{Fano,Bethe}.

\section{An improvement of the  M--F theory}

\subsection{Glauber approximation}

The  multiple scattering amplitude can be represented in the Glauber
approximation \cite{Glauber} as
 \ba\label{gl1} F_{if}(\vec q)=\frac{ik}{2\pi} \int d^2\rho\, \exp(i\,\vec q\vec
b)\,\Gamma_{if}(\vec \rho), \ea where  $\Gamma_{if}(\vec \rho)$ is
so-called `profile function'.

We can get a general formulation of the problem by considering the
scattering of a pointlike projectile on a system of $Z$ constituents
with the coordinates $\vec r_1,\,\vec r_2,\ldots\,,\vec
r_{\scriptscriptstyle Z}$ and the projections on the plane of the
impact parameter $\vec s_1,\,\vec s_2,\ldots\,,\vec
s_{\scriptscriptstyle Z}$. Then the total phase shift can be written
as a sum of the form
\ba \widetilde\chi(\vec \rho,\vec x_1,\,\vec
x_2,\ldots\,,\vec x_{\scriptscriptstyle
Z})=\sum\limits_{k=1}^{Z}\widetilde\chi_k(\vec \rho-\vec s_k). \ea
If we introduce the configuration space for the wave functions
$\psi_i$ and $\psi_f$ in the initial $i$ and the final $f$
constituent's states, the profile function can be presented as
\ba\label{Gamma}\Gamma_{if}(\vec \rho)&=& \int \prod_{k=1}^{Z}d^3r_k
\,\psi_f^{\ast}(\{\vec r_k\})\psi_i(\{\vec r_k\})\Gamma(\vec
\rho,\{\vec s_k\})\ea
with an interaction operator
\ba \Gamma(\vec \rho,\{\vec s_k\})=1- \exp[i\Phi(\vec \rho,\{\vec
s_k\})] \ea
and a phase-shift function
\ba\label{Phi} \widetilde{\Phi}(\vec \rho,\{\vec s_k\})= Z\widetilde\chi(\vec
\rho)-\sum_{k=1}^{Z}\widetilde\chi(\vec \rho-\vec s_k).\ea
When the interaction is due to a potential $V(\vec r)$, the phase
function $\widetilde\chi(\vec \rho)$ is given by
\ba \label{chi}\widetilde\chi(\vec \rho)&=&-\,\frac{1}{v}
\int\limits_{-\infty}^{\infty}V\left(\sqrt{\rho^2+z^2}\right)dz \ea
with  the potential of an individual constituent's

\ba\label{gln} V(r)=\,\pm \lim\limits_{\lambda\rightarrow
0}\frac{\alpha}{r}\, e^{-\lambda r},\quad \lambda\sim m_e\alpha
Z^{1/3}. \ea

The multiple-scattering amplitude $F_{if}(\vec q)$ \eqref{gl1} is
normalized by the relations
\ba 4\pi\Im F_{ii}(0)=\sigma(i)_{tot}, \quad\vert F_{if}(\vec
q)\vert^2=d\sigma_{if}/dq_{\scriptscriptstyle T},\ea where \ba
\sigma(i)_{tot}=\sigma(i)_{el}+ \sigma(i)_{in},\quad
\sigma_{if}=\int \vert d\sigma_{if}/dq_{\scriptscriptstyle T}\vert^2
d^2q ,\ea\ba\sigma(i)_{tot}=\sum_{f}\sigma_{if}. \ea

In terms of $e^{i\widetilde{\Phi}}$, where the phase-shift function
$\widetilde{\Phi}=\widetilde{\Phi}(\vec \rho,\{\vec s_k\})$ is given
by \eqref{Phi}, the cross sections $\sigma(i)_{tot}$,
$\sigma(i)_{el}$, and $\sigma(i)_{in}$ become
\ba \sigma(i)_{tot}=2\Re\int\Big\langle 1-\left\langle
e^{i\widetilde{\Phi}} \right\rangle \Big\rangle \,d^2\rho, \ea
\ba \sigma(i)_{el}=\int\Big\langle 1-2\Re\left\langle
e^{i\widetilde{\Phi}} \right\rangle +\left\vert\left\langle
e^{i\widetilde{\Phi}}\right\rangle \right\vert^2\Big\rangle \,d^2\rho,
\ea
\ba \sigma(i)_{in}=\int\Big\langle 1-\left\vert\left\langle
e^{i\widetilde{\Phi}} \right\rangle\right\vert^2\Big\rangle \,d^2\rho.
\ea
The brackets $\left\langle e^{i\widetilde{\Phi}}\right\rangle$
signify that averaging is performed over all the configurations of
the target constituents' in $i$th state.

\subsection{Reconstruction of unitarity conditions}

To reduce the above many-body problem to the consideration of an
effective one-body one and to establish the relationship between the
Glauber and M--F theories, we introduce an abbreviation
\ba\langle e^{i\widetilde{\Phi}}\rangle=e^{i\overline{\Phi}}.\ea
For the effective (`optical') phase shift function
$\overline{\Phi}(\vec \rho)$, we will consider the following
expansion
\ba\label{expan} \overline{\Phi}(\vec
\rho)=\sum_{n=1}^{\infty}\frac{i^{n-1}}{n!}\Phi_n,\ea
 where
$$ \Phi_1=\left\langle \Phi\right\rangle,\quad \Phi_2=\left\langle
(\Phi-\Phi_1)^2\right\rangle,$$ \ba\Phi_3= \left\langle
(\Phi-\Phi_1)^3 \right\rangle,\ldots , \Phi_n\sim Z\alpha^n/\beta.
\ea

The first order for $\overline{\Phi}(\vec \rho)$ is simply the
average of the function $\widetilde{\Phi}(\vec \rho,\{\vec s_k\})$;
it correspond to the first-order Born approximation. The
second-order term of $\overline{\Phi}(\vec \rho)$ is purely
absorptive; it is equal in order of magnitude to $Z\alpha^2/\beta$.

When the remainder term $\overline{\Phi}_{3}(\vec \rho)$ in the
series \eqref{expan} is much smaller than unity
 \ba \overline{\Phi}_{3}(\vec
\rho)=\sum_{n=3}^{\infty}\frac{i^{n-1}}{n!}\Phi_n\ll 1,\ea
it seems natural to neglect them and consider the following
approximation:
\ba \overline{\Phi}(\vec \rho)\approx \Phi_1(\vec
\rho)+\frac{i}{2}\Phi_2(\vec \rho),\ea
in which we put $\Phi_1(\vec \rho)=\Phi_{\scriptscriptstyle M}(\vec
\rho)$ and $\Phi_2(\vec \rho)=2\Phi_{in}(\vec \rho)$. The last term
corresponds to the target-inelastic (incoherent) scattering.
%, also
%referred to as incoherent scattering or `antiscreening'
%\cite{Basel}.

This leads to the following improvement of the Moli\`ere--Fano
theory:
\ba \Phi_{\scriptscriptstyle M}^{}(\vec \rho)\Rightarrow
\Phi_{\scriptscriptstyle M}^{}(\vec \rho)+i\Phi_{in}(\vec \rho)\ea
with
\ba 2\Phi_{in}(\vec \rho)=\lim\limits_{\lambda\rightarrow 0}
Z\left\{\int\left\vert\widetilde\chi^{}_{\lambda}(\vec \rho -\vec
r_{\scriptscriptstyle T})\right\vert^2\varrho(\vec
r)d^3r-\left\vert\int\widetilde\chi^{}_{\lambda}(\vec \rho -\vec
r_{\scriptscriptstyle T})\varrho(\vec
r)d^3r\right\vert^2\right\},\ea
where
\ba \widetilde\chi^{}_{\lambda}(\vec
\rho)=\,-\,\frac{1}{v}\int\limits_{-\infty}^{\infty}V^{}_{\lambda}
\left(\sqrt{\rho^2+z^2}\right)dz, \quad\,
V^{}_{\lambda}(r)=\mp\frac{\alpha }{r}e^{-\lambda r},\ea
\ba\varrho(\vec r)=\psi_f^{\ast}(\vec r)\psi_i(\vec r).\ea

For the cross sections
 \ba
\sigma_{tot}=\left\langle\sigma(i)_{tot}\right\rangle,\quad
\sigma_{in}=\left\langle\sigma(i)_{in}\right\rangle,\quad
\sigma_{el}=\left\langle\sigma(i)_{el}\right\rangle,\ea
the following unitarity condition is valid:
\ba
\label{uk}\Im f_{el}(0)=\frac{k}{4\pi}\sigma_{tot}
%_{tot}
=\frac{k}{4\pi}(\sigma_{el}+ \sigma_{in}) \ea
with
\ba f_{el}(\theta)=F_{ii}(\vec q),\quad \Im F_{ii}(0)=\frac{k}{4\pi}
\sum_{f}\int\left\vert F_{if}(\vec q)\right\vert^2d\Omega,\ea
\ba \frac{d\sigma_{in}}{d\Omega}= \sum_{f\neq i}\vert F_{if}(\vec
q)\vert^2,\quad F_{if}(\vec q)=\frac{ik}{2\pi} \int d^2\rho\,
\exp(i\vec q\vec \rho)\,\Gamma_{if}(\vec \rho), \ea \ba
\Gamma_{if}(\vec \rho)=1-\exp(-2\Phi_{in}), \ea
$$q=k\theta.$$
Making use of \eqref{uk}, we can find the following expressions for
the cross sections $\sigma_{tot}$, $\sigma_{in}$, and $\sigma_{el}$:
\ba \sigma_{tot}=4\pi \int \left(1-\cos\Phi_{\scriptscriptstyle
M}^{}(\vec \rho)\,e^{-\Phi_{in}(\vec \rho)}\right)\rho\,d\rho,\ea
\ba \sigma_{in}=2\pi \int \left(1-e^{-2\Phi_{in}(\vec
\rho)}\right)\rho\,d\rho,\ea
\ba \sigma_{el}=2\pi \int \left(1-2\cos\Phi_{\scriptscriptstyle
M}^{}(\vec \rho)\,e^{-\Phi_{in}(\vec \rho)}+e^{-2\Phi_{in}(\vec
\rho)}\right)\rho\,d\rho.\ea

\subsection{Unitarity corrections to the Born approximation}

Using the evaluation formula
\ba\int [2\Phi_{in}(\vec \rho\,)]\,d^2\rho\sim Z\alpha^2 /\beta \ea
and the exact contributions have been calculated in \cite{tarvoskr},
we obtain the following unitarity relative correction
($\delta_{\scriptscriptstyle UN}\equiv \delta_{\scriptscriptstyle
UN}^{(2)}$) to the first-order Born cross section of the inelastic
scattering $\sigma_{in}^{B}$:
\ba \label{cor2}\delta_{\scriptscriptstyle
UN}=\frac{\Delta\,\sigma_{in}}{\sigma_{in}^{B}}
=\frac{\sigma_{in}-\sigma_{in}^{B}}{
\sigma_{in}^{B}}=\frac{\sigma_{in}}{\sigma_{in}^B}-1\sim
Z\alpha^2/\beta \ea
with
\ba\sigma_{in}^{B}=\left\langle\int\Phi_{in}(\vec \rho)d^2\rho
\right\rangle.\ea
The corresponding angular distribution reads
\ba\label{W_in} W_{in} (\theta)=2\pi \int\limits_0^{\infty}
J_0(\theta \eta)e^{-Q_{in} (\eta)} \eta\, d\eta, \ea \ba\label{Q_in}
Q_{in} (\eta)=2\pi \int \sigma_{in} (\theta)[1-J_0(\theta
\eta)]\theta d\theta. \ea
Inserting \eqref{Q_in} back into \eqref{W_in}, we get the equation
of the form:
\ba\label{com} W_{in} (\theta)=2\pi \int\limits_0^{\infty}\eta\,
d\eta \,J_0(\theta \eta\,)\exp\left[-2\pi \int
\limits_0^{\infty}\sigma_{in}(\theta^{\,\prime})\theta^{\,\prime}
d\theta^{\,\prime}[1-J_0(\theta^{\,\prime} \eta\,)]\right]. \ea
With the use of
\ba\label{delta}\int \eta\, J_0(\theta\eta) J_0(\theta^{\,\prime}
\eta\,)d\eta =\frac{1}{\theta}\,\delta(\theta-\theta^{\prime})=0 \ea
%for $\theta\neq 0$,
and
\ba\label{Gamm} \int\limits_0^{\infty}d\eta\,\eta\,
J_{0}(\theta\eta)=2a^{-2}\,\frac{\Gamma\left(1\right)}{\Gamma\left(0\right)}=0,\quad
\int\limits_0^{\infty}d\eta\,\eta\,
J_{0}(\theta\eta)J_{0}(\theta^{\prime}\eta)=0, \ea according to
\cite{ryzh}, the integration of \eqref{com} yields the following
result:
\ba\label{sol} W_{in} (\theta)=-(2\pi)^2
\int\limits_0^{\infty}\eta\, J_0(\,\theta\eta\,)
J_0(\theta^{\,\prime} \eta\,)d\eta
\cdot\sigma_{in}(\theta^{\,\prime})\theta^{\,\prime}
d\theta^{\,\prime} =-(2\pi)^2\,\sigma_{in}(\theta).\ea
In \eqref{delta} and \eqref{Gamm}, $\delta$ is the Dirac delta
function, and $\Gamma$ is the Euler Gamma function.

Finally, taking into account the relations \eqref{cor2} and
\eqref{sol}, we can estimate the unitarity correction to the angular
distribution function \eqref{W}:
\ba\label{unit} \delta_{\scriptscriptstyle UN}=\frac{\Delta
W_{in}(\theta)}{W_{in}^B(\theta)}= \frac{W_{in}(\theta)
}{W_{in}^B(\theta)}-1 =\frac{\sigma_{in}(\theta)
}{\sigma_{in}^B(\theta)}-1=\frac{\sigma_{in} }{\sigma_{in}^B}-1 \sim
Z\alpha^2/\beta.\ea

\subsection{Coulomb corrections to the Born approximation results}
%Rigorous relations between the exact and Born results}

Recently, it has been shown within the eikonal approach \cite{prd}
by means of \eqref{gl1}--\eqref{gln} that the following rigorous
relation between the quantities $\ln\big[g(\eta)\big]$ and
$\ln\big[g^{\scriptscriptstyle B}(\eta)\big]$ holds:
\ba\label{g} \Delta\big[\!-\ln
g(\eta)\big]&=&\Delta_{\scriptscriptstyle CC}\left[\ln
g(\eta)\right]=\frac{1}{2}\left(\chi_c\eta\right)^2\Delta_{\scriptscriptstyle
CC}\left[\ln\big(\chi_a^{\,\prime}\big)^2\right]\nonumber\\
&=&\left(\chi_c\eta\right)^2\frac{1}{2\pi}\int d^2x \left[
\left(\frac{(\vec{x}+
\vec{b})^2}{x^{\,2}}\right)^{i\xi}-1+\frac{a^2}{2}
\ln^2\frac{(\vec{x} +\vec{ b})^2}{x^{\,2}} \right]
%\begin{equation}
=\left(\chi_c\eta\right)^2f(a) \ea
%nd{equation}
%
with the Coulomb corrections $\Delta_{\scriptscriptstyle CC}\big[\ln
g(\eta)\big]\equiv \ln g(\eta)-\ln g^{\scriptscriptstyle B}(\eta)$,
$\Delta_{\scriptscriptstyle
CC}[\ln\big(\chi_a^{\,\prime}\big)]\equiv\ln\big(\chi_a^{\,\prime}\big)-\ln
\big(\chi_a^{\,\prime}\big)^{\scriptscriptstyle B}$, and
$\chi_a^{\,\prime}\equiv 1.080\,\chi_a$. Here, $\vec x=\gamma\vec
b$, $\gamma$ is the usual relativistic factor of the scattered
particle, $\vec b$ is impact-parameter vector, and $f(a)$ is an
universal function of the Born parameter $a=Z\alpha/\beta$, which is
also known as the Bethe--Maximon function:
\begin{equation}\label{summa} f(a)=a^2\sum_{n=1}^\infty\frac{1}{n(n^2+a^2)}\ .
\end{equation}
This universal function can be evaluated  by means of the expression
\cite{ryzh} (see details in \cite{revis}):
\ba f(a)
&=&1-\frac{1}{1+a^2}+0.2021\,a^2-0.0369\,a^4+0.0083\,a^6-\ldots\label{f(a)}
\ea

The above exact analytical expression \eqref{g} immediately leads to
the corresponding result for the screening angle
\ba \label{basres1} \Delta_{\scriptscriptstyle
CC}[\ln\big(\chi_a^{\,\prime}\big)]=f(a). \ea
For the specified value of $\eta^2=1/\chi^2_c$, \eqref{g}  also
becomes this form
\ba\label{result3} \Delta_{\scriptscriptstyle CC}[ \ln g(\chi_c)]=
\frac{4\pi n_0t}{\chi^2_c}\frac{\chi_c^2}{4\pi \,n_0t}\, f(a)=f(a).
\ea
Thus,
\ba\label{positiv}\Delta_{\scriptscriptstyle
CC}\left[\ln\big(\chi_a^{\,\prime}\big)\right]=\Delta_{\scriptscriptstyle
CC}\left[\ln g(\chi_c)\right]=f(a)>0. \ea

For the relative Coulomb correction to the Born screening angle
$\big(\chi_a^{\,\prime}\big)^{\scriptscriptstyle B}$, we get
\ba\label{del0}\delta_{\scriptscriptstyle
CC}\big(\chi_a^{\,\prime}\big)=\frac{\chi_a^{\,\prime}-
\big(\chi_a^{\,\prime}\big)^{\scriptscriptstyle B}}
{\big(\chi_a^{\,\prime}\big)^{\scriptscriptstyle
B}}=\frac{\Delta_{\scriptscriptstyle CC}\big(\chi_a^{\,\prime}\big)}
{~~~~\big(\chi_a^{\,\prime}\big)^{\scriptscriptstyle B}}=
\frac{\Delta_{\scriptscriptstyle CC}\big(\chi_a^{}\big)}
{\chi_a^{\scriptscriptstyle B}} =\exp
\left[f\left(Z\alpha/\beta\right)\right]-1. \ea\\
The relative CC to the Bessel-transformed distribution function
$g^{\scriptscriptstyle B}(\eta)$ can also be determined by this
quantity for $\eta^2=1/\chi^2_c$:
$$\delta_{\scriptscriptstyle
CC}\big(\chi_a^{\,\prime}\big)=\delta_{\scriptscriptstyle
CC}\big(\chi_a^{}\big)=\delta_{\scriptscriptstyle
CC}\left[g(\chi_c)\right].$$
Besides, because
\ba\label{mend} \Delta_{\scriptscriptstyle CC}[W(\chi_c,t)]\equiv
W_{\scriptscriptstyle M}- W^{\scriptscriptstyle
B}_{\scriptscriptstyle M}= \int\limits_0^{\infty} J_0(\theta
\eta)\Delta_{\scriptscriptstyle CC}[g(\chi_c)] \eta d\eta, \ea
accounting for $\int\nolimits_0^{\infty}d\eta\,\eta\,
J_{0}(\theta\eta)=0$, we arrive at the following result:
\ba\label{mendW} \delta_{\scriptscriptstyle
CC}\left[W_{\scriptscriptstyle
M}(\chi_c,t)\right]=\frac{\Delta_{\scriptscriptstyle CC}
[W(\chi_c,t)]} {W^{\scriptscriptstyle
B}(\chi_c,t)_{\scriptscriptstyle
M}}=\frac{\Delta_{\scriptscriptstyle CC}[g(\chi_c)]}
{g^{\scriptscriptstyle B}(\chi_c)} =\exp
\left[f\left(Z\alpha/\beta\right)\right]-1. \ea
Consequently,
\ba\label{univ}\delta_{\scriptscriptstyle
CC}\equiv\delta_{\scriptscriptstyle CC}\big(\chi_a^{\,\prime}\big)=
\delta_{\scriptscriptstyle CC}\left[g(\chi_c)\right]=
\delta_{\scriptscriptstyle CC}\left[W_{\scriptscriptstyle
M}(\chi_c,t)\right]=\exp[f(a)]-1>0.\ea

The Coulomb corrections to the parameters of the Moli\`{e}re
expansion method, i.e. $b$, $B$, and $\overline{\theta^2}$,
according to \cite{prd}, are as follows:
\ba\label{limcorrec1} \Delta_{\scriptscriptstyle
CC}(b)&=&-f(a),\\
\label{limco1} \Delta_{\scriptscriptstyle CC}(B)&=&\frac{f(a)}{1/B^{\scriptscriptstyle B}-1}\ ,\\
\label{CCvarthet2} \Delta_{\scriptscriptstyle
CC}\left(\overline{\theta^2}\right)&=&\chi_c^2
\cdot\Delta_{\scriptscriptstyle CC}\left(B\right)\ , \ea
and the relative Coulomb corrections become
\begin{equation}\label{rel}
\delta_{\scriptscriptstyle
CC}\left(\overline{\theta^2}\right)=\delta_{\scriptscriptstyle
CC}\left(B\right) =\frac{f(a)}{1-B^{\scriptscriptstyle B}}\ .
\end{equation}

In contrast to the unusual positive value of the above Coulomb
corrections \eqref{positiv} and \eqref{univ}, these  Coulomb
corrections \eqref{limcorrec1}--\eqref{rel} have a negative value.
Furthermore, as can be seen from \eqref{limco1} and \eqref{rel},
these Coulomb corrections are dependent on the
$B^{\scriptscriptstyle B}$ value. This dependence is presented in
Table 1 for some separate sizes of $Z$ over the entire range $4.5
\leq B^{\scriptscriptstyle B}\leq 20$ of  the parameter
$B^{\scriptscriptstyle B}$, which provide the convergence of the
expansion series \eqref{ex}.

\bigskip

\begin{center}
{\bf Table 1.}  The dependence of the Coulomb corrections
$\Delta_{\scriptscriptstyle CC}\left(B\right)$ (\ref{limco1}) and
$\delta_{\scriptscriptstyle CC}\left(\overline{\theta^2}\right)$
(\ref{rel}) on the $B^{\scriptscriptstyle B}$ value over the range
$4.5 \leq B^{\scriptscriptstyle B}\leq 20.5$

\bigskip

{\bf 1.} for $Z=92$ and $f(Z\alpha)=0.3951$

\medskip

\begin{tabular}{lccccccccc}
\hline \\[-3mm]
$B^{\scriptscriptstyle B}$&4.5&6.5 &8.5 &10.5 &12.5&14.5&16.5&18.5&20.5\\[.2cm]
\hline\\[-3mm]
$-\Delta_{\scriptscriptstyle CC}\left(B\right)$&
0.5080&0.4669&0.4478&0.4367&0.4295&0.4244&0.4206&0.4177&0.4154\\
$-\delta_{\scriptscriptstyle CC}\left(\overline{\theta^2}\right)$&
0.1129&0.0718&0.0527&0.0416&0.0344&0.0293&0.0254&0.0226&0.0203\\[.2cm]
\hline
\end{tabular}

\bigskip

{\bf 2.} for $Z=79$ and $f(Z\alpha)=0.3125$

\medskip

\begin{tabular}{lccccccccc}
\hline \\[-3mm]
$B^{\scriptscriptstyle B}$&4.5&6.5 &8.5 &10.5 &12.5&14.5&16.5&18.5&20.5\\[.2cm]
\hline\\[-3mm]
$-\Delta_{\scriptscriptstyle CC}\left(B\right)$&
0.4080&0.3693&0.3542&0.3454&0.3397&0.3357&0.3327&0.3304&0.3285\\
$-\delta_{\scriptscriptstyle CC}\left(\overline{\theta^2}\right)$&
0.0893&0.0568&0.0417&0.0329&0.0272&0.0231&0.0202&0.0179&0.0160\\[.2cm]
\hline
\end{tabular}

\bigskip

{\bf 3.} for $Z=50$ and $f(Z\alpha)=0.1436$

\medskip

\begin{tabular}{lccccccccc}
\hline \\[-3mm]
$B^{\scriptscriptstyle B}$&4.5&6.5 &8.5 &10.5 &12.5&14.5&16.5&18.5&20.5\\[.2cm]
\hline\\[-3mm]
$-\Delta_{\scriptscriptstyle CC}\left(B\right)$&
0.1846&0.1697&0.1628&0.1587&0.1561&0.1542&0.1529&0.1518&0.1510\\
$-\delta_{\scriptscriptstyle CC}\left(\overline{\theta^2}\right)$&
0.0410&0.0261&0.0191&0.0151&0.0125&0.0106&0.0093&0.0082&0.0074\\[.2cm]
\hline
\end{tabular}

\bigskip

{\bf 3.} for $Z=28$ and $f(Z\alpha)=0.0487$

\medskip

\begin{tabular}{lccccccccc}
\hline \\[-3mm]
$B^{\scriptscriptstyle B}$&4.5&6.5 &8.5 &10.5 &12.5&14.5&16.5&18.5&20.5\\[.2cm]
\hline\\[-3mm]
$-\Delta_{\scriptscriptstyle CC}\left(B\right)$&
0.0626&0.0575&0.0552&0.0538&0.0529&0.0523&0.0518&0.0515&0.0512\\
$-\delta_{\scriptscriptstyle CC}\left(\overline{\theta^2}\right)$&
0.0139&0.0088&0.0065&0.0051&0.0042&0.0036&0.0031&0.0028&0.0025\\[.2cm]
\hline
\end{tabular}

\end{center}

\bigskip

It demonstrates that modules of these corrections decrease with
increasing $B^{\scriptscriptstyle B}$ and $Z$. For target material
of the DIRAC experiment ($Z=28$), these corrections are negligible.
However, in conditions of the SLAC experiment \cite{slac} (the gold
and uranium targets, and $B^{\scriptscriptstyle B}=8.46$ ), the
above Coulomb corrections are essential.

\newpage

\begin{center}

{\bf Table 2.} The $Z$ dependence of the corrections and differences
defined by \eqref{unit}, \eqref{f(a)}, and
\eqref{univ}--\eqref{angle} for $B^{\scriptscriptstyle B}=4.5$.

\end{center}

\begin{center}

\begin{tabular}{rccccccccc}
\hline\\[-2mm]
$Z$~~~&$~10\,\delta_{\scriptscriptstyle UN}~$&$f(Z\alpha)$&
$\delta_{\scriptscriptstyle CC}(\chi_a)$&$\delta_{\scriptscriptstyle
M}(\chi_a)$&$\delta_{\scriptscriptstyle
CCM}(\delta_{\scriptscriptstyle CC})$& $\delta_{\scriptscriptstyle
CCM}(\chi_a)$&$\delta_{\scriptscriptstyle
CC}\!\big(\overline{\theta^2}\big)$&$\Delta_{\scriptscriptstyle
CC}\left(b\right)$&$\Delta_{\scriptscriptstyle CC}\left(B\right)$\\
\hline\\[-2mm]
4~~~& 0.002&0.001~~~&0.001&0.001&$-0.299$&$-0.000$&$-0.000$&$-0.001$&$-0.001$\\
13~~~&0.007&0.011~~~&0.011&0.015&$-0.276$&$-0.004$&$-0.003$&$-0.011$&$-0.014$\\
22~~~&0.012&0.030~~~&0.031&0.042&$-0.270$&$-0.011$&$-0.009$&$-0.030$&$-0.039$\\
28~~~&0.015&0.049~~~&0.050&0.068&$-0.265$&$-0.017$&$-0.014$&$-0.049$&$-0.063$\\
42~~~&0.022&0.105~~~&0.110&0.146&$-0.246$&$-0.031$&$-0.030$&$-0.105$&$-0.134$\\
50~~~&0.027&0.144~~~&0.154&0.202&$-0.235$&$-0.040$&$-0.041$&$-0.144$&$-0.185$\\
73~~~&0.039&0.276~~~&0.318&0.396&$-0.198$&$-0.056$&$-0.079$&$-0.276$&$-0.355$\\
78~~~&0.041&0.307~~~&0.359&0.443&$-0.189$&$-0.058$&$-0.088$&$-0.307$&$-0.394$\\
79~~~&0.042&0.312~~~&0.367&0.452&$-0.188$&$-0.059$&$-0.089$&$-0.312$&$-0.402$\\
82~~~&0.044&0.332~~~&0.393&0.482&$-0.185$&$-0.060$&$-0.095$&$-0.332$&$-0.426$\\
92~~~&0.050&0.395~~~&0.484&0.583&$-0.169$&$-0.062$&$-0.113$&$-0.395$&$-0.508$\\[.2cm]
\hline
\end{tabular}

\end{center}

\bigskip

Also, we estimate the accuracy of the Moli\`{e}re theory in
determining the Coulomb correction to the screening angle
$\chi_a^{}$ by means of the difference
 and relative difference between the values of
$\delta^{}_{\scriptscriptstyle M}\big(\chi_a^{}\big)$ and
$\delta_{\scriptscriptstyle CC}\big(\chi_a^{}\big)$:
\ba\label{corr3}\delta_{\scriptscriptstyle
CCM}(\delta_{\scriptscriptstyle
CC})=\frac{\Delta_{\scriptscriptstyle
CCM}(\delta_{\scriptscriptstyle CC})}{\delta^{}_{\scriptscriptstyle
M}\big(\chi_a^{}\big)} =-\frac{\delta^{}_{\scriptscriptstyle
CC}\big(\chi_a^{}\big)-\delta_{\scriptscriptstyle
M}\big(\chi_a^{}\big)}{\delta^{}_{\scriptscriptstyle
M}\big(\chi_a^{}\big)}, \ea
where
\ba\label{mol} \delta^{}_{\scriptscriptstyle
M}\big(\chi_a^{}\big)=\frac{\chi^{}_a-\chi_a^{\scriptscriptstyle
B}}{\chi_a^{\scriptscriptstyle B}}=\sqrt{1+3.34}-1. \ea

The accuracy of the Moli\`{e}re theory in determining the screening
angle is estimated by the following relative difference between the
approximate $\chi_a^{\scriptscriptstyle M}$ and exact $\chi_a$
results
\begin{eqnarray}\label{angle}
\delta_{\scriptscriptstyle CCM}(\chi_a)&\equiv&\frac{\chi_a-
\chi_a^{\scriptscriptstyle M}}{\chi_a^{\scriptscriptstyle M}}=
\frac{\Delta_{\scriptscriptstyle CCM}(\delta_{\scriptscriptstyle
CC})}{\delta^{}_{\scriptscriptstyle M}(\chi_a)+1}.
%\label{rat}
\end{eqnarray}

The calculation results for the unitarity and Coulomb  corrections,
as well as the differences between our results and those of
Moli\`{e}re over the range $4\leq Z \leq 92$ are presented in Table
2. Some results from Table 2 are represented by Figure 1.

The Table 2 shows that while the value of relative unitarity
correction $\delta_{\scriptscriptstyle
UN}\equiv\delta_{\scriptscriptstyle
UN}(\sigma)=\delta_{\scriptscriptstyle UN}(W)$ reach only $0.5\%$
for heavy atoms of the target material, the maximum value of the
relative Coulomb correction $\delta_{\scriptscriptstyle
CC}\equiv\delta_{\scriptscriptstyle CC}\big(\chi_a^{\,\prime}\big)=
\delta_{\scriptscriptstyle CC}\left[g(\chi_c)\right]=
\delta_{\scriptscriptstyle CC}\left[W_{\scriptscriptstyle
M}(\chi_c,t)\right]$ is two orders of magnitude higher and amounts
approximately to 50\% for $Z=92$ (Fig. 1).

From Table 2 and Figure 1 it is also obvious that the difference
$\Delta_{\scriptscriptstyle CCM}\left(\delta_{\scriptscriptstyle
CC}\right)$ between our results and those of Moli\`{e}re in
determining the relative Coulomb correction to the screening angle
increases to $10\%$ with the rise of $Z$, and the corresponding
relative difference $\delta_{\scriptscriptstyle
CCM}\left(\delta_{\scriptscriptstyle CC}\right)$ varies between 30
and 17\% over the range $4\leq Z \leq 92$. The
$\delta_{\scriptscriptstyle CCM}\left(\chi_a\right)$  value amounts
about to $6\%$ for $Z=80\div 90$.

The modules of the CC to the parameters $b^{\scriptscriptstyle B}$
and $B^{\scriptscriptstyle B}$ reach large values for high Z
targets. For instance, $-\Delta_{\scriptscriptstyle CC}(B)\sim 0.50$
and $-\Delta_{\scriptscriptstyle CC}(b)\sim 0.40$ for
$B^{\scriptscriptstyle B}=4.5$ and $Z=92$. The modulus of  Coulomb
correction to the mean square scattering angle is about $11\%$ for
$Z=92$. This needs to be accounted for, if one aims a quantitative
interpretation of experimental data.

Thus, the such  large Coulomb corrections as
$\Delta_{\scriptscriptstyle CC}\equiv\Delta_{\scriptscriptstyle
CC}\left[\ln\big(\chi_a^{\,\prime}\big)\right]=\Delta_{\scriptscriptstyle
CC}\left[\ln g(\chi_c)\right]=f(a)$, $\delta_{\scriptscriptstyle
CC}\equiv\delta_{\scriptscriptstyle CC}\big(\chi_a^{\,\prime}\big)=
\delta_{\scriptscriptstyle CC}\left[g(\chi_c)\right]=
\delta_{\scriptscriptstyle CC}\left[W_{\scriptscriptstyle
M}(\chi_c,t)\right]$, $-\Delta_{\scriptscriptstyle CC}(B)$, and
$-\Delta_{\scriptscriptstyle CC}(b)$ should be taken into account in
experimental data analysis. The accuracy of the Moli\`{e}re theory
in determining the Coulomb correction to the screening angle must
also be taken into consideration for a rather accurate description
of the experimental data.

%in which the relevant correlations become small, whereas Coulomb
%corrections become crucial to extract information about the
%structure functions and to understand the Coulomb sum rule. 28

\newpage

\begin{figure}[h!]

\begin{center}

\includegraphics[width=0.67\linewidth]{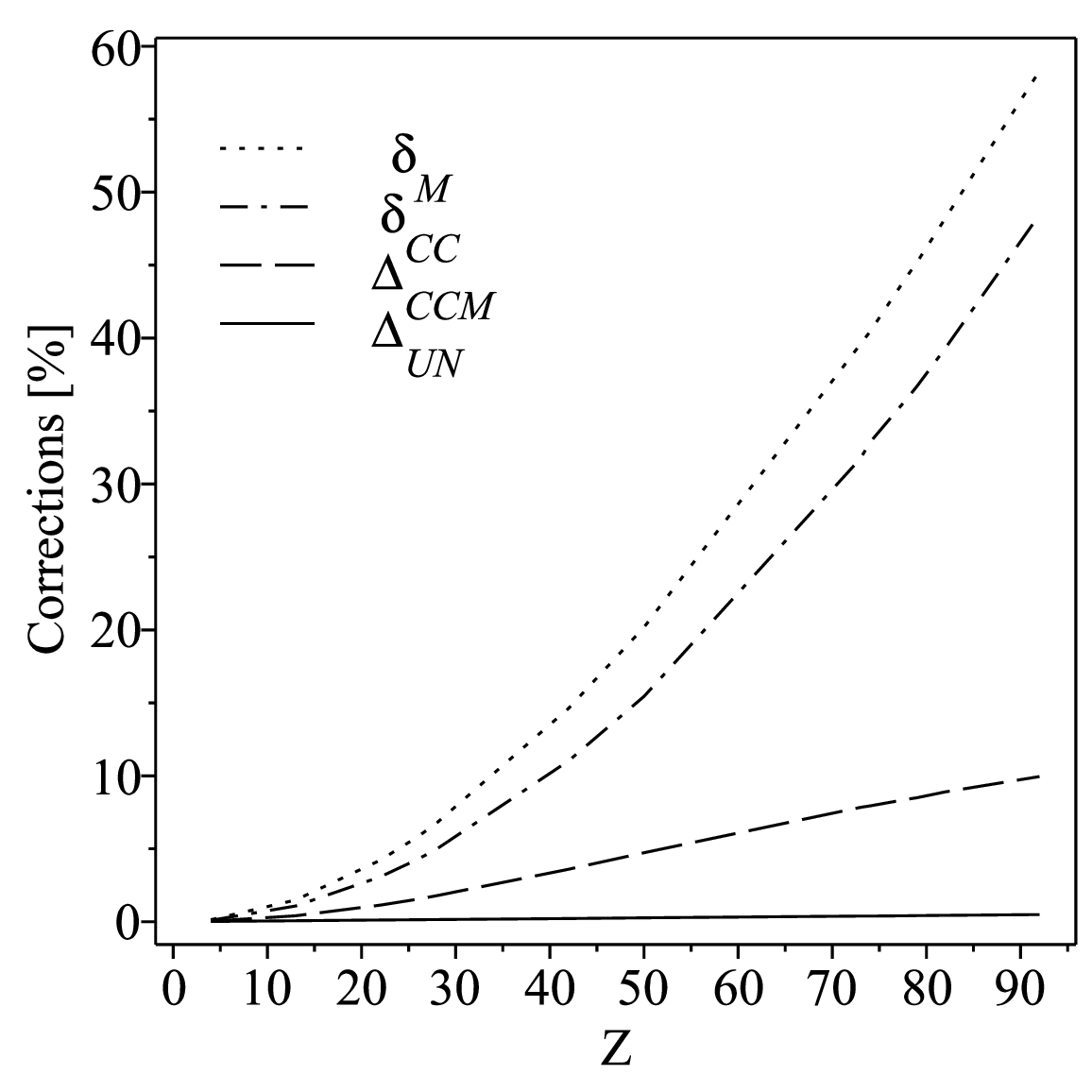}

\caption{\small The dependence of the relative Coulomb
($\delta_{\scriptscriptstyle CC}$) and Moli\`{e}re
($\delta^{}_{\scriptscriptstyle M}$) corrections to the screening
angle, their difference ($\Delta_{\scriptscriptstyle CCM}$), and the
unitarity correction ($\delta_{\scriptscriptstyle UN}$) on the
nuclear charge $Z$.} \label{Fig1}

\end{center}

\end{figure}

\section{Summary}

1. Within the framework of fully unitary Glauber approximation for
particle-atom scattering, we develop the general formalism of the
Moli\`ere--Fano multiple scattering theory.

2. We have estimated the relative unitarity correction
$\delta_{\scriptscriptstyle UN}\equiv\delta_{\scriptscriptstyle
UN}(\sigma)=\delta_{\scriptscriptstyle UN}(W)$ to some quantities of
the M--F theory resulting from reconstruction of its unitarity in
the second-order optical model of the Glauber theory, and we found
that they are of an order of $Z\alpha^2$.

3. Within the eikonal approach, we have considered rigorous
relations between the exact and Born values of the quantities
$\ln\left[ g(\eta)\right]$ and $\ln\big(\chi_a^{\,\prime}\big)$.
Also, we calculated the Coulomb corrections
$\Delta_{\scriptscriptstyle CC}\equiv\Delta_{\scriptscriptstyle
CC}\left[\ln\big(\chi_a^{\,\prime}\big)\right]=\Delta_{\scriptscriptstyle
CC}\left[\ln g(\chi_c)\right]$ and relative Coulomb corrections
$\delta_{\scriptscriptstyle CC}\equiv\delta_{\scriptscriptstyle
CC}\big(\chi_a^{\,\prime}\big)= \delta_{\scriptscriptstyle
CC}\left[g(\chi_c)\right]= \delta_{\scriptscriptstyle
CC}\left[W_{\scriptscriptstyle M}(\chi_c,t)\right]$ for nuclear
charge ranged from $Z=4$ to $Z=92$, and  we showed that these
corrections
%comprise the order
increase up to $40$ and $50\%$, correspondingly, for $Z=92$.

4. Besides, we have obtained analytical and numerical results for
the Coulomb corrections to the parameters of the Moli\`{e}re
expansion method ($b^{\scriptscriptstyle B}$, $B^{\scriptscriptstyle
B}$, and $\big(\overline{\theta^2}\big)^{\scriptscriptstyle B}$),
which depend on the sizes of $B^{\scriptscriptstyle B}$ and Z.  We
have examined their $B^{\scriptscriptstyle B}$ and $Z$ dependences
over the ranges $4.5 \leq B^{\scriptscriptstyle B}\leq 20.5$ and $4
\leq Z\leq 92$, and found that while the correction
$\delta_{\scriptscriptstyle CC}\!\big(\overline{\theta^2}\big)$
becomes the value about 11\%, the corrections
$-\Delta_{\scriptscriptstyle CC}[B]\times 10^2$ and
$-\Delta_{\scriptscriptstyle CC}[b]\times 10^2 $ become very large
value (about $40\div 50$) at small $B^{\scriptscriptstyle B}$ and
large $Z$.

5. Additionally, we have evaluated the inaccuracies of the
Moli\`{e}re theory in determining the relative Coulomb correction to
the screening angle. We shoved that its absolute inaccuracy
$\Delta_{\scriptscriptstyle CCM}$ reach about $10\%$ for $Z=92$, and
the corresponding relative inaccuracy $\delta_{\scriptscriptstyle
CCM}$ varies between 30 and 17\% over the range $4\leq Z \leq 92$.

\section*{Appendix: Derivation of the transport equation for the Bessel-transformed
distribution function}
%% \label{}

We put here the details of inferring Eq. (\ref{eq5}). We apply first
the integration operation $\int\nolimits_0^\infty \theta d\theta
J_0(\eta\theta)$ to both sides of (\ref{W}). Using the definition of
the Bessel transform of the probability distribution (\ref{eq3}), we
obtain
\ba\label{A1} \frac{\partial g(\eta,t)}{\partial
t}=-n^{}_0\,g(\eta,t)\int\nolimits_0^\infty \sigma^{}_{el}(\chi)\chi
d\chi+ n^{}_0\int\nolimits_0^\infty \sigma^{}_{el}(\chi)\chi d\chi
I(\eta,\chi) \ea
with
\ba I(\eta,\chi)=\int\limits_0^{2\pi}\frac{d\phi}{2\pi}\int \theta
d\theta J_0(\eta\theta) W_{\scriptscriptstyle
M}\left(\left\vert\vec{\chi}-\vec{\theta}\right\vert,t\right)\ . \ea
Applying the opposite Bessel transform to the probability (\ref{W}),
we get for the last integral
\ba I(\eta,\chi)=\int\limits_0^{2\pi}\frac{d\phi}{2\pi}\int \theta
d\theta J_0(\eta\theta)\int\limits_0^\infty \eta_1 d\eta_1
J_0\left(\eta_1\left\vert\vec{\chi}-\vec{\theta}\right\vert\right)g(\eta_1,t)\
,\ea
where the integration over $\theta$ can be performed using the
folding theorem
\ba \label{A4}\int\limits_0^{2\pi}\frac{d\phi}{2\pi}
J_0\left(\eta_1\left\vert\vec{\chi}-\vec{\theta}\right\vert\right)=
J_0(\eta_1\theta)J_0(\eta_1\chi)\ . \ea
With the means of the orthogonality relation for the Bessel
functions
\ba \int\limits_0^\infty x d x J_0(x a)J_0(x
b)-\frac{1}{a}\delta(a-b), \ea
we get for $I(\eta,\chi)$:
\ba\label{A6} I(\eta,\chi)=g(\eta,t)J_0(\eta\chi)\ . \ea
Inserting (\ref{A6}) into (\ref{A1}), we immediately arrive at a
result:
$$\frac{\partial
g(\eta,t)}{\partial t}=-n^{}_0\,
g(\eta,t)\int\limits_0^\infty\sigma^{}_{el}(\chi) \chi
d\chi[1-J_0(\eta\chi)]\  .$$
To prove the folding theorem (\ref{A4}), we use the series expansion
for the Bessel function
 \ba
 J_0(z)=1-\frac{(z^2/4)}{(1!)^2}+\frac{(z^2/4)^2}{(2!)^2}-
 \ldots,
 %\;\;
\ea
\ba z^2=\eta^2\left[\theta^2+\chi^2-2\theta\chi\cos\phi\right]
  \ea
and perform the integration over $\phi$:
\ba\frac{1}{2\pi}\int\limits_0^{2\pi}(\cos\phi)^{2n}d\phi=\frac{(2n-1)!!}{(2n)!!}\
.\ea

\end{document}